# On Thermal Sensation and Entropy

Douglas M. Snyder
Los Angeles, California

The definition of the temperature of a physical system involves the entropy, and thus degree of disorder, in that system. Temperature relates change in the entropy of a system to change in the energy of that system. How is it that an individual can experience the temperature of objects in the physical world when disordered processes are part of the foundation of temperature? How are disordered processes accounted for in an individual's cognition when this cognition is concerned with finding order in the world? What are the neurophysiological concomitants to thermal experience that account for the disordered processes fundamental to temperature? This last question is the subject of this paper.

More precisely, the temperature of a physical system may be defined as $1/T = k_B (\partial \sigma / \partial U)_A$ where T is the temperature given in Kelvin, $\sigma$ is the entropy of the component system, U is the energy of the system, $k_B$ is the Boltzmann constant, and A represents any characteristics of the system other than energy that affect the system's entropy (Kittel, 1969). The symbol $\partial$ indicates partial differentiation and thus the term $(\partial \sigma / \partial U)_A$ indicates the partial differentiation of the entropy with respect to the energy, while other characteristics of the system that affect the entropy are held constant. Examples of these characteristics are the number of particles in the system and the volume of the system. How can disordered processes that are central to the temperature of a physical object be accounted for on a neurophysiological level in thermal sensation?

Perhaps the most famous discussion in modern Western philosophy of the experience of temperature, and its underlying physiological concomitants, is given by Locke (1690/1959) in discussing secondary qualities in the perception of physical objects:

> We may be able to give an account how the same water, at the same time, may produce the idea of cold by one hand and of heat by the other: whereas it is impossible that the same water, if those ideas were really in it, should at the same time be both hot and cold. For, if we imagine *warmth*, as it is in our hands, to be nothing but a certain sort and degree of motion in the minute particles of our nerves or animal spirits, we may understand

>how it is possible that the same water may, at the same time, produce the sensations of heat in one hand and cold in the other; which yet *figure* never does, that never producing the idea of a square by one hand which has produced the idea of a globe by another. But if the sensation of heat and cold be nothing but the increase or diminution of the motion of the minute parts of our bodies, caused by the corpuscles of any other body, it is easy to be understood, that if that motion be greater in one hand than in the other; if a body be applied to the two hands, which has in its minute particles a greater motion than in those of one of the hands, and a less than in those of the other, it will increase the motion of the one hand and lessen it in the other; and so cause the different sensations of heat and cold that depend thereon. (p. 177)

Locke, of course, did not know statistical mechanics, which was developed about 100 years ago and in which the modern definition of temperature is rooted. Locke did not know that disordered processes essential to temperature would need to be accounted for in his attribution of the experience of temperature to the "sort and degree of motion in the minute particles of our nerves or animal spirits" (p. 177). Locke thought that thermal sensation is based on energy alone in that the sensation of different temperatures is due to different amounts of motion at the periphery of the body.

Science has found that thermal sensitivity is mediated by a number of different neuronal receptors (Martin & Jessel, 1991). These receptors are dorsal root ganglion neurons with cell bodies located near the spinal cord. Among these receptors are those dedicated to sensing temperatures less than a person's adapted skin temperature (termed "cold" receptors), temperatures greater than a person's adapted skin temperature (termed "warm" receptors), receptors that serve to sense significant heat in association with pain (a type of nociceptor), and receptors that serve to sense significant cold in association with pain (also a type of nociceptor). The peripheral terminal of all of these neurons is a bare nerve ending. Cold receptors are activated in a range from about $1^{o}C$ to $20^{o}C$ below normal skin temperature, approximately $34^{o}C$. Warm receptors are activated in the range of about $32^{o}C$ to $45^{o}C$. Above about $45^{o}C$, the discharge of warm receptors decreases, and the nociceptors associated with the sensation of heat and pain respond.

# On Thermal Sensation

The characteristic of the warm and cold receptors that is of concern here is the positive correlation that has been empirically established between human subjective experience of thermal stimuli and the rate of discharge of the warm and cold receptors in monkey over a large part of the range in which they are activated (Martin & Jessel, 1991). (Locke was correct in apparently deducing that the experience of warmth and cold is to a significant extent determined at the periphery.) And as Darian-Smith (1984) wrote in his review article on thermal sensation, "cutaneous thermal stimuli are presented with considerable fidelity, not only in the peripheral fiber responses but also in the central neural processes concerned with the subjective response" (p. 908). Thus, scientific evidence supports our personal experience that each of us in general has a significant capacity to accurately distinguish the temperature of external physical objects.

Allowing for Locke's concern regarding the relative character of thermal scaling in our experience, the question remains: How does the experience of temperature of external physical objects occur when the temperature of these objects in the physical world involves fundamentally disordered processes? What are the neurophysiological concomitants that allow for the thermal sensation of physical objects? Darian-Smith (1984) wrote in the *Handbook of Physiology* that the transductive process that initiates thermoreceptive processes was not well-known and that the nature of the processing related to thermal sensation in the central nervous system also was not well-known. It is proposed that at some level in the neurophysiological processing of thermal stimuli, the *disordered processes* fundamental to temperature are accounted for. There is likely some form of thermal equilibrium rapidly established between thermoreceptors and the external environment local to those thermoreceptors. Though an approximation, the physical object being sensed and the receptor attain a state of thermal equilibrium in accordance with the equation describing thermal equilibrium for two physical systems that together comprise an isolated physical system:

$$k_B (\partial \sigma_1 / \partial U_1)_{A_1} = k_B (\partial \sigma_2 / \partial U_2)_{A_2}$$

or

$$T_1 = T_2$$

where the subscript 1 indicates values of quantities in one of the component systems (e.g., the external physical object) and the subscript 2 indicates values of quantities in the other component system (e.g., the receptor) (Kittel, 1969).

# On Thermal Sensation

Information received at the receptor level in the establishment of thermal equilibrium is then transmitted accurately to higher level processing stations. It is interesting that in the neurophysiological processes involved in thermal sensation, information that fundamentally relies on disordered processes at the receptor level is accurately transmitted in the form of ordered processes as action potentials along the peripheral neurons involved in thermal sensation. Just as the operation of a furnace may depend on the disordered processes of the sensitive element in the thermostat, so the conduction of the action potentials in the peripheral neurons and in higher levels of the nervous system related to thermal sensation may depend as well on the disordered processes involved in the flow of thermal energy between receptor endings of peripheral neurons and the physical object being thermally sensed. This energy flow is approximately in accordance with the tendency to establish thermal equilibrium for systems in thermal contact which together constitute an isolated system. Indeed, the tendency to establish thermal equilibrium for these systems is due to the increase in the entropy of the overall system that occurs when systems move toward thermal equilibrium.

The net energy change in the receptor in contact with the physical object may be negative. That is, if the temperature of the object is lower than the temperature of the receptor, energy will flow from the receptor to the physical object in the establishment of thermal equilibrium.

The scenario outlined should hold over the temperature range where thermal sensitivity is precise. That is where thermal equilibrium should be quickly established between an external process and a thermoreceptor. Where the accuracy of thermal sensitivity is substantially decreased, one would not expect the rapid establishment of thermal equilibrium between thermoreceptors and external processes.